\documentclass[11pt]{article}
\usepackage{graphicx}
\usepackage[margin=1.25in]{geometry}
\usepackage[usenames,dvipsnames]{color}
\usepackage{url}
\usepackage[colorlinks = true,
            linkcolor = blue,
            urlcolor  = blue,
            citecolor = blue,
            anchorcolor = blue]{hyperref}

\newenvironment{Abstract}{\begin{quotation} \begin{center}
                       ABSTRACT
     \end{center}\bigskip  }{\end{quotation}}

\def\Acknowledgements{\bigskip  \bigskip \begin{center} \begin{large}
             \bf ACKNOWLEDGEMENTS \end{large}\end{center}}

\def\Acknowledgements{\bigskip  \bigskip \begin{center} \begin{large}
             \bf ACKNOWLEDGEMENTS \end{large}\end{center}}

\def\beq{\begin{equation}}
\def\eeq#1{\label{#1}\end{equation}}
\def\eeqn{\end{equation}}

\newenvironment{Eqnarray}%
   {\arraycolsep 0.14em\begin{eqnarray}}{\end{eqnarray}}
\def\beqa{\begin{Eqnarray}}
\def\eeqa#1{\label{#1}\end{Eqnarray}}
\def\eeqan{\end{Eqnarray}}

\let\bar=\overbar

\def\lsim{\mathrel{\raise.3ex\hbox{$<$\kern-.75em\lower1ex\hbox{$\sim$}}}}
\def\gsim{\mathrel{\raise.3ex\hbox{$>$\kern-.75em\lower1ex\hbox{$\sim$}}}}

\def\del{\partial}
\def\Dslash{\not{\hbox{\kern-4pt $D$}}}
\def\dslash{\not{\hbox{\kern-2pt $\del$}}}
\def\pslash{\not{\hbox{\kern-2pt $p$}}}
\def\ETmiss{\not{\hbox{\kern-4pt $E$}}_T}

\def\Dlr{\mathrel{\raise1.5ex\hbox{$\leftrightarrow$\kern-1em\lower1.5ex\hbox{$D$}}}}

\def\MSB{{\bar{M \kern -2pt S}}}
\def\msb{{\bar{\scriptsize M \kern -1pt S}}}

\def\drb{{\bar{\scriptsize D \kern -1pt R}}}

\newcommand\snowmass{\begin{center}\rule[-0.2in]{\hsize}{0.01in}\\\rule{\hsize}{0.01in}\\
\vskip 0.1in Submitted to the  Proceedings of the US Community Study\\ 
on the Future of Particle Physics (Snowmass 2021)\\ 
\vspace{0.2cm}
\it{Snowmass 2021 CEF03 Diversity, Equity \& Inclusion}
\rule{\hsize}{0.01in}\\\rule[+0.2in]{\hsize}{0.01in} \end{center}}

\usepackage{authblk}
\title{\huge{Lifestyle and personal wellness in particle physics
research activities}}

\author[1,2]{Tiffany R. Lewis}
\author[3]{Sara M. Simon}
\author[4,5]{Carla Bonifazi}
\author[6]{Savannah Thais}
\author[7]{Johan Sebastian Bonilla Castro}
\author[8]{K\'et\'evi A. Assamagan}
\author[9]{Thomas Y. Chen}

\affil[1]{Astroparticle Physics Laboratory, NASA Goddard Space Flight Center, Greenbelt, MD, USA}
\affil[2]{NASA Postdoctoral Program Fellow}
\affil[3]{Fermi National Accelerator Laboratory, Batavia, IL, USA}
\affil[4]{ICAS – ICIFI, ECyT-UNSAM
and CONICET, Argentina}
\affil[5] {Universidade Federal do Rio de Janeiro, Brazil}
\affil[6]{Princeton University}
\affil[7]{University of California, Davis, USA}
\affil[8]{Brookhaven National Laboratory, Physics Department, Upton, New York, USA}
\affil[9]{Columbia University}

\begin{document}


\maketitle

\medskip

 \begin{Abstract}
\noindent 
Finding a balance between professional responsibilities and personal priorities is a great challenge of contemporary life and particularly within the HEPAC community. Failure to achieve a proper balance often leads to different degrees of mental and physical issues and affects work performance. In this paper, we discuss some of the main causes that lead to the imbalance between work and personal life in our academic field. We present some recommendations in order to establish mechanisms to create a healthier and more equitable work environment, for the different members of our community at the different levels of their careers.

\end{Abstract}

\snowmass

\def\thefootnote{\fnsymbol{footnote}}
\setcounter{footnote}{0}
\newpage
\tableofcontents
\newpage
\section{Introduction}

Our community is immersed in a very competitive work environment, which generates considerable stress in the search for a balance between work and other personal needs. This is particularly true for Early Career members, who commonly feel the need to stand out as competitive and successful researchers to ensure future employment and resources. The competition in particle physics is particularly stressful as often people working together in the same collaboration may end up competing for the same job positions. Very often individuals wind up working after hours and during the weekends, without respecting their needs for health, leisure, vacation, physical activities, etc. This can cause long-term health problems due to the stress, lack of physical activity, and lack of psychological support, which, in turn, can lead to depression, anxiety, and other serious illness.

These stressful work practices typically have a larger effect on those wishing to follow a career in the academic world than those seeking less precarious jobs in physics. The demands of an academic career are for many more akin to the rat race often associated with careers in finance, insurance, and technology. 

Having a collective understanding of the importance to set and respect the boundaries between work and personal life in the field of HEPAC is needed to define a new framework for collaborative work. Industries and institutions that put people first perform the best in the long run because they draw in and retain the best minds. These practices have yet to be widely implemented in the academic world.

In this contribution, we discuss some root causes of the competition and the imbalances of current academic life in HEPAC and provide a set of recommendations to alleviate some of these problems. We also discuss the impact of the COVID-19 pandemic, which has often exacerbated the imbalances already present in academic life and has disproportionately affected women, gender minorities, and underrepresented groups in the field.

\section{Causes of work-life imbalance}

The root causes of work-life imbalance are systemic and far-reaching. Societal expectations around care roles, the competitive work environment in academia, bias in evaluation metrics, unclear and incomplete job expectations, low pay and limited access to resources, and discrimination all contribute to these issues. In this section, we discuss these sources of work-life imbalance present in the academic field in detail. We note that these are some primary causes, but note that this discussion does not include all sources of imbalance.

\subsection{Gender Unbalanced Care Roles \& Impact on Compensation}

The proper role of an academic institution in the life of its academic persons is to best facilitate their work. Since life and culture deal unequal hands to equally capable minds, it falls to institutions and collaborations to reduce undue burdens to liberate time, mental energy, and efficiency for the people it invests money and resources in. This argument is not counter to the interests of those institutions, but rather a road map for how to achieve peak scientific output from a finite number of people over decades. 

The United States is one of only two countries worldwide with no guaranteed leave for new parents. As such, it falls to individual companies, organizations and institutions to create a policy for parental leave. Parental leave is critically important both for persons giving birth to physically recover and for all new parents to adjust to this major life event. Giving birth is a significant physical process on par with a major surgery, and it can often involve a hospital stay or surgical intervention. Policies around leave should reflect that significance. This physical consideration must include cases of stillbirth for the same reason. Since physics academia has historically been male dominated, especially in positions with sufficient power to set such policies, there has been a lack of understanding surrounding the gravity of the event and how to account for it from a managerial perspective. 

In addition to the physical act of giving birth, the presence of a new infant in the home is a unique, major life event. Parental leave policies must include adoptive parents. Access to resources to assist with the care of a newborn is highly sensitive to race, ethnicity, socioeconomic background, and family history. Additionally, the cultural expectation that women are solely responsible for child-rearing, especially for very young children, often makes parental leave extremely gendered. Parental leave affects this transitional phase for the child and anyone else residing in the home. The lack of adequate leave during this time often leads to people leaving academia for jobs that do offer leave. In particular, women and other underrepresented groups (with emphasis on intersectionality) who do not feel they have sufficient practical support will are more likely to leave the field. Therefore, parental leave policy has a significant impact on people's ability to stay in the field.

The impact of inadequate parental leave is nearly impossible to measure in surveys of people who stay in the field because those are not the people for whom this type of policy was an insurmountable or distasteful barrier. Family planning also tends to be a private topic that individuals may not feel comfortable discussing with their professional colleagues, employer, potential employers, or a survey that will get back to the aforementioned persons. Some policies regarding personnel need to be designed to anticipate common scenarios. While the birth or adoption of a child may be a rare event in the life of an individual, it is sufficiently common on a population level that it should be anticipated by employers. 

Another reason for parental leave is to cater to the unique needs of the child during the time of transition into the home. This is true of infants born to their parents as well as adoptions or recent placements (e.g. foster care, family emergency) of infants and older children in the home. In considering this broader set of circumstances, it is not appropriate here to define the timeline upon which each child or circumstance ought to be accommodated by the parental employer, except that whatever policy the employer settles on should be flexible enough to take into account individualized recommendations from the relevant professionals involved. Those professionals in the case of an infant might be the pediatrician or in the case of adoption a therapist or social worker assigned to the case. 

Other care responsibilities also often fall unequally to women, including care for older children, disabled family, and elders. Since these tasks tend to be highly gendered, it is important to challenge the status quo and to systematically encourage balancing the load between parents, of which overwhelmingly tends to be carried by female-identifying parents. Thus, it is important to offer and normalize taking family leave to all academics and to foster a culture of taking time for all genders of employees. While it is often the case in academia that people are hired due to their unique abilities and training, the burden of carrying out necessary, time-sensitive tasks should never fall to one person alone. Therefore, appropriate hiring policy such that the burdens of the institution are not placed on an individual is an important aspect of institutional planning. 

Students are often not considered in conversations surrounding parental leave, but it is important to recognize that some people will have children as students and universities and departments should not leave parental leave policies to individual advisors. The department policies for pregnant persons and new parents should be similar across academic rank and age. While staffing in smaller departments can be a challenge, the burden of staffing shortages should not fall on pregnant persons and new parents. It is the responsibility of the admissions and hiring committees to make sure that there are enough people to perform necessary work with the understanding that medical events should be expected to arise. 

Furthermore, it should go without saying, but since there are places in physics it doesn't, we state clearly here: \textit{The potential for a person to have a child should never be a reason not to hire them or not to admit them to an academic program.} The pay of women and gender minorities should not be reduced in anticipation of the time they might spend on parental leave or on activities outside of their contracted duties/hours. 

In order to set more uniform standards across the field (lacking federal legislation or any specific expectation for it), there is a role for collaborations and academic societies in setting minimum standards for member institutions and job ads they are willing to distribute. 

A reason that some families choose to rely on male income when one person is needed at home is that men tend to bring in more money. If all genders were compensated according to their skills and job functions, without influence by their perceived station in life, then it might be less common for women and gender minorities to leave careers. And, women and gender minorities from lower socioeconomic backgrounds might find a career in physics to be more acceptable if they received equitable training, compensation, and sense of stability in their jobs.

Single parents and persons from lower socioeconomic backgrounds do not have the choice of not working, and working without job security is especially stressful. While we often talk about the choices women and gender minorities make surrounding career vs. homemaker, that is not and has never been the choice many women and gender minorities are presented with. That dichotomy is one that comes with a level of financial security that many women and gender minorities do not have. The choice may be something more like academia (flexible, but less reliable employment, long hours, relatively low pay) or industry (less flexible, but more permanent employment, fixed hours, relatively higher pay). It is not difficult to see why someone might chose to leave the field under those circumstances. 

Cultural expectations of gender roles can impact the career choices of women and gender minorities, but they are not predictive of intelligence or capability. Women and gender minorities who choose to adhere to specific styles of dress or habits of grooming might contend with arbitrary rules or biases that should simply be removed from codes of conduct where present and specifically allowed where they need to be acknowledged. Some cultures hold expectations of parenting to effectively be a bias against admitting or hiring certain intersectional identities due to gendered religious or cultural bias. Women and gender minorities who flout such cultural expectations and thereby give up acceptance in their cultural or familial community and require an additional level of support from their career and academic peers.

\subsection{Very competitive work environment}

As in physics in general, the field of HEPAC is very competitive. Most students completing their PhD struggle to continue in the academy and only a fraction is able to obtain post-doc positions on the mainstream research universities and national laboratories. The path to a tenured position is long and has also a high number of candidates per position. This competition inevitably leads to a work/personal life imbalance, where working after hours, during holidays, etc., has become the rule. The pressure on productivity often makes students risky averse, so that they may end up taking less original projects. The emphasis on productivity may also decrease the overall quality of the work due to the rush to publish, leaving less time for improvements, cross checks, careful reading of the literature, etc. The competition for positions has also a negative effect on the collaboration among researchers and students.

In the case of experimental research on HEPAC, there is a specificity due to the fact that a large fraction of the work is carried out in large collaborations, involving a growing number of researchers, institutions and countries. Research on particle colliders has involved large international collaborations for several decades. More recently, a growing fraction of the research in astrophysics and cosmology has also been carried out by large collaborations and same holds for neutrino and dark matter detection, among other fields. As the experiments in these fields have been growing in complexity and cost, the same holds for the size of the collaborations, which increased from several dozens of researchers to several hundreds or even a few thousand.

Despite the size of the collaborations and although the work is distributed within each collaboration, most experiments are still person power limited and, furthermore, the load is unevenly distributed, such that many members are overloaded. This particularly affects early career researchers, who have to show intense activity, fight to take important roles in the experiment, and stand out among other colleagues to advance to their next position. Although the size of the collaborations are growing, the number of the associate positions in universities and research laboratories has not increased at the same pace, so positions are even more competitive.

There is also the aspect of uneven rewards and recognition regarding different roles in the experiment. In many cases, being directly involved in the construction of the experiment or in other more technical activities, such as software development or experiment operation, pays off less in terms of future employability. Being directly involved in data analysis, especially in conjunction with theoretical and phenomenological work entitles more recognition, particularly when viewed from outside the collaboration. This also produces an imbalance in the work load, as early career scientists tend to choose activities that have higher visibility and recognition.

In some research fields, the large size of the experiment and collaboration is such that it encompasses a large fraction of the researchers in a given field. Therefore, instead of competition among experiments, there is internal competition among scientists within the same collaboration, which contributes to an unhealthy work culture.


\subsection{Bias in Evaluation Metrics}

Evaluation for admission or employment should be based on well thought out, prepared rubrics. In the case of continued employment or tenure evaluation, those rubrics should be known to the employee ahead of being evaluated by them. Committees who create such rubrics should consider carefully and attempt to research if any of the criteria are biased in a way that would limit access by people who are traditionally underrepresented. 

Examples of common pitfalls in seemingly unbiased criteria: 

Different expectations of women and gender minorities from students and mentees. Even if the employment environment for women and gender minorities is the same from the top down, the expectations from students and mentees can be vastly different. College students tend to expect professors that are women and gender minorities to be more nurturing and have lower expectations than male professors. They can become combative and critical impacting both the classroom environment and eventually the course/teaching reviews. Student evaluations of instructors that are women and gender minorities at the college level tend to be more negative than those of male instructors in part because of a lack of respect for expertise and in part because of the additional expectation of nurturing that may be inappropriate in the context of a college course. Evaluations of teaching performance for the purpose of continued employment or tenure can thus be impacted by the gendered biases of students. Departments have a responsibility for taking this known bias into account if they choose to include student feedback in employment or tenure evaluations. Similar effects may also be present for other axes of minority and should be expected to be of exacerbated impact for instructors with intersectional identities. 

When designing an undergraduate physics program that will serve a diverse set of students it is important to consider both what the students should be expected to achieve by the end of the program, and the preparation they should be expected to have prior to their arrival. In particular it is quite common for undergraduate physics programs to assume that any student who is worth teaching physics to will have mastered at least one semester's worth of calculus during high school. Programs that are built with this assumption deter some students from entering the field by forcing students to choose between finishing a degree in 4 years or attempting to study physics. It is not necessary to create this false exclusivity to maintain high standards in the individual courses or in the program overall. However, it will negatively impact the diversity of the program. 

Less than 20\% of high school students in the United States take calculus prior to graduation \footnote{https://www.nsf.gov/statistics/2018/nsb20181/report/sections/elementary-and-secondary-mathematics-and-science-education/high-school-coursetaking-in-mathematics-and-science}.  While calculus achievement is nearly equal between genders, the differences across racial and socioeconomic divides are stark. Additionally, about 40\% of high school graduates are sufficiently prepared to take Calculus in their first semester in college, meaning not only that the number of high school graduates who could in principle major in physics would be tripled by structuring the program to allow for first semester calculus, but the racial and socioeconomic barriers to entry into a physics degree would be significantly diminished by this single change. A student's race or socioeconomic background is not indicative of their intelligence or ability to thrive in a well structured physics program. 

Activities that aim to center the human aspects of recruiting scientists, managing the public image of science, and improving the working conditions of scientists through the identification of inappropriate barriers to research participation and the countering of them are often considered passion projects rather than integral parts of long-term maintenance and planning for the field. Private companies pay teams of people to perform these kinds of tasks in public image management, human resource development and community building. While many physicists view it as a moral imperative to participate in building up these aspects of the field, it is rare that such activities are considered to be part of the job and therefore exceedingly rare that time spent on such activities results in compensation or recognition for the time spent. Work in outreach; equity, diversity, and inclusion (EDI) initiatives; and non-research mentoring should be paid for by the employers, collaborations, and grant programs that it benefits in addition to being recognized as important service to the community in consideration for academic jobs, as evidence of dedication to the betterment of the community and the field. 

The work of mentoring, outreach, and diversity initiatives falls unevenly on minoritized groups. It is no more the responsibility of scientists from underrepresented groups to improve the field for people who look like them than it is for any other member of the community. The goal of increasing diversity is not served by overburdening a token individual with all work related to underrepresented groups. Specifically in the realm of mentoring, advisors who belong to underrepresented groups in the field tend to receive more requests for more of their time from mentees, increasing the burden of mentoring students on a small number of people. Institutions and departments should be aware of this kind of trend and have measures in place either to more evenly distribute student mentoring requests or to make accommodations for mentors who spend extra time serving the needs of their students. While it may be important for some students to see ``someone who looks like them" in a career they aspire to, it is not appropriate to foster a system that pairs students with advisors based on their respective demographics. 

In evaluating scientists, there are some easily accessible, numerical metrics that are often used. This is acceptable as a part of a holistic evaluation of a candidate for a job or seniority, but it is important to consider the inherent bias in any external metric when comparing candidates. For example, it is common to look at the number of publications a candidate has and how many publications for which they were first author. Especially for broad solicitations, it is important to keep in mind that different subfields have vastly different publication cultures. Additionally, persons associated with a collaboration or not, working in theory vs. instrumentation, can all have very different understandings of publication frequency, who should write papers, who should have their name added to a paper they didn't write and what order authors should appear in. So, in evaluating candidates, it is important to evaluate their publication record in the context in which they produced it, which may be quite different from the context in which someone else produced a very different publication record. 

It has been shown that women and gender minorities are extremely underrepresented as PIs on grant applications as well as in grant proposal teams. While it is likely that other underrepresented groups are also lacking in grant team and especially PI representation, the discrepancies are nearly impossible to account for when demographic information for individual proposals is not collected as a matter of course. Some initiatives to improve this situation, or even to monitor it include collecting PI and team demographic information on grant proposals, and making that aggregate information publicly available and using dual-anonymous peer review practices to evaluate proposals.  Additionally, the NASA ROSES Astrophysics Theory Program recently released the results of a pilot program that asked all proposers to include a statement on how their work would serve diversity, equity and inclusion in the field, and that statement was evaluated by a different review panel comprised of experts in STEM EDI and social science.  It is important that access to funding be an equitable process and that scientific prospects are judged on the merits of individual proposals so that the best science is funded and carried out.

There are power dynamics at play at every stage of a physicist's career, but in practice, one of the most problematic can be between a PhD candidate and their advisor because the student tends to be less experienced in navigating interpersonal challenges in a professional environment and are less aware of of collegial standards in the field. It is more difficult to challenge an authority figure when one is unsure of whether their behavior is objectively problematic or not and to what degree. Additionally letters of recommendation from PhD advisors are often required in future job applications, so challenging an advisor could put future job prospects in jeopardy. The solutions to this problem are threefold. (1) All students and most postdocs should have a second advisor or mentor either at their institution or within their collaboration. Studies demonstrate that women who have 2 advisors in graduate school are more likely to finish their program. It is likely that this trend would also bear out for other underrepresented and intersectional identifying students as well as postdocs.  A second advisor gives the mentee a second opinion on research topics, an additional repository of skills and expertise from which to learn, and an additional anchor in the broader community. (2) Departments should implement a system of faculty overseeing each other's advising. This can take the form of a designated faculty member checking in with students regularly and being available for questions. More commonly, it takes the form of a faculty committee that connects with students in different research areas as a group to socialize and keep track of how students are doing without the burden of judging those students on their work at any point. The Fermi Collaboration has implemented a version of this system externally, where students are paired with non-research mentors to support them in their career development without the students having to worry about sharing their struggles with research since the mentors are not involved. (3) Departments and Collaborations should take initiative to train students and postdocs specifically in the standards of the field. Early career scientists who know what they are expected to do at various stages of their careers and how they should be treated are better prepared to make informed decisions about who they work with and the work they want to do in the future. While much of this role is traditionally held by advisors and advisors should still attend to individualizing the information, these types of programs are a hedge against any unconscious bias or gatekeeping of career information that some students may face as an inappropriate hurdle to their desired career trajectory.

Some physics instructors grade students according to their perception of student understanding without or in spite of grading actual work. This can sometimes arise as a coping mechanism for classes that are larger than the instructor is prepared to grade for properly, but the solution is to find a grader or reduce the instructor's other responsibilities in compensation for the additional workload. Grading based on perception or even in-class participation tends to reward confidence moreso than competence. Any system that operates as a reward for confidence alone will be subject to 2 major fallacies. (1) People who are better able to identify their weaknesses tend to be more competent and less confident, especially early in their careers. (2) People who feel the burden of their underrepresented identities in the context of a physics classroom tend to be less outwardly confident because of a fear of being judged more harshly if they make a mistake that should be expected in a learning environment. So, making judgments about a student's capabilities based on the confidence in their responses will tend to result in worse science outcomes and problems in retaining a diverse cohort.

In designing demographic surveys for departments or institutions, it is best practice to involve a social scientist as an expert throughout the process. This comes with a cost as the social scientist and/or their students will need to be compensated for their time and resources. Grants for the purpose should be made available to scientific programs that might have need of them. Institutions should also have a vested interest in providing funding to make sure that their educational and work environments are serving the needs of all of their students and academic staff. The reason to include a social scientist's expertise in any survey is that they have specific training in how to perform surveys without bias, obtain the best results, and to properly analyze the results while maintaining anonymity and respect for the participants and their responses. When it is not possible to include the expertise of a social scientist, it becomes more important to avoid as much bias as possible through the survey committee. It would be ideal in that case for some physicists to be trained in EDI work and surveying specifically - this seems like something that could be offered as a targeted certificate program as more institutions want scientists to be actively involved in EDI work. The next best approach probably involves forming a committee with a diverse set of backgrounds, such that as many axes of representation as possible are accounted for (gender, orientation, career stage, race/ethnicity, socioeconomic background, etc). 

While the first thing many departments want to do when beginning EDI initiatives is to perform a survey, it is imperative to the goal of not doing harm that there be a plan in place for what to do with the results. Will the results be published or kept confidential? Are there resources and political will available to make changes within the institution if the survey indicates that changes are needed? Has the committee formulated specific plans of action for each possible outcome? If any of these questions are not answered in the affirmative, then the survey is not ready to be released. It is unethical to use ones colleagues and students as test subjects in a diversity experiment.

\subsection{Work time}
The physics community is global. This can often lead to meetings and correspondence occurring at times that fall out of nominal work hours in a specific time zone. While efforts are made to find times for meetings that can work in multiple time zones, it is not always possible, especially when trying to accommodate many time zones. There is also added pressure to respond to messages received from those working in different time zones outside of work hours.

\subsection{Job Expectations}
Job expectations are often not clearly stated, and there are often many unspoken expectations. Research and/or teaching duties are often made clear and have clear funding streams, but other duties often fall outside of funding. Service work like equity, diversity, and inclusion, outreach, and mentoring often falls into this category. Additionally, activities like journal refereeing, grant and project reviews, and hiring committees are often unfunded. Since funding typically only covers research and teaching duties, there is pressure to do these additional duties outside of work hours since they are unpaid or to not invest time in them at all. Underrepresented groups in physics are often oversubscribed with service work, so this often also leads to disparities in who is shouldering this unpaid and unacknowledged work. These duties are critical to the scientific community, so they need to be adequately funded so that they can be part of the nominal work day.

\subsection{Low Pay and Limited Access to Resources until Tenure-track}

In addition to a lack of pay for key parts of the job, early stages in academic careers have low salaries and few resources. Pay for postdocs and graduate students is often well below what they could make in industry. This can lead to early career members seeking additional sources of income to support themselves and/or their families. The low pay for these positions can create socioeconomic barriers. For example, a student from a disadvantaged socioeconomic background may not be able to take a low paying graduate school position because they may need to help support their families or have other financial obligations (e.g. student loan debt).

The financial implications extend beyond individual pay. Some fellowships are paid as educational funds, preventing students from making retirement contributions, putting them at a long-term financial disadvantage. Other fellowships count salary, stipend, and tuition fully as taxable income, so taxes can be for much more income than graduate students receive as pay. These are enormous financial barriers that many universities do not prioritize. 

Additionally, most academic positions to not offer relocation services for individuals or their partners until tenure-track positions if at all. Often those in academic positions need to move to a new job every few years. The cost of moving can be steep, especially for international moves. Sometimes subsidized housing is available through universities, but it often excludes partners and/or spouses. Moving locations often can also affect the careers of partners and spouses. It often necessitates a change in jobs for their partners, but little to no support is given in finding partners/spouses job placements when relocating until the tenure-track level if at all. Additionally, a long pattern of moving can make a partner or spouse's resume look suspect to potential employers. Health care is also a barrier when relocating for research. Often spouses and children are covered near the campus of one's employer, but if the researcher is stationed abroad for research, institutions often have no precedent, guidelines, or willingness to cover additional health care costs abroad.

\subsection{Harassment and Discrimination}

Many forms of harassment and discrimination are illegal in the United States. Most universities and national labs have some official method of reporting inappropriate conduct.   Students, researchers and academics often interact more with their collaboration than their home institution, where reporting avenues are less available. It is expected that large societies and meeting hosts have a code of conduct for members, especially surrounding national conferences. Reports of harassment and discrimination given to representatives of an institution (e.g. Title IX coordinators \& Ombudspersons) tend to produce mixed results for the complainant, while reports given to designated `ethics officers' (which may go by many names) at conferences tend to produce positive outcomes for the complainant. Better results in terms of support offered to the complainant, if not a sense of justice have been reported in complainant interactions with persons of authority who do not perceive their role as representing the interests of the institution. Ideas about reportable harassment and discrimination are discussed further in the {\it Snowmass CEF03 white paper on ``Climate of the Field"}. 

A person whose mental energies are imposed upon by the biases of their colleagues, students, or superiors is less likely to perform optimally in their role as a scientist. It is the responsibility of the employer to create a working culture that supports the mental health and well-being of all of the employees they chose to invest in by hiring. Mental energies are certainly sapped by macro-aggressions like assault, sexual coercion, and blatant racism. Without diminishing the impact of that kind of harassment and discrimination, or the role of current power structures and assessment criterion in perpetuating them, it is also important to recognize that less overt biases also impact the mental health and productivity students, researchers and academics. These less overt biases may be expressed as stereotyping, exclusion, sarcastic comments about individuals morals or capabilities, or as an adherence to poorly defined ``cultural fit."  The specifics take many forms and are often collectively referred to as ``micro-aggressions" or the kind of subtle insult to ones morals or intelligence that cannot be politely refuted and yet weighs unequally on the the minds of targeted individuals building up over time and repetition to a hostile work environment. 

The immediate evidence of microaggressions in action can look like self-fulfilling stereotyping, for example if a professor perceives a student to be from a less privileged background, they might assume that the student is less capable and therefore recommend the student pursue coursework and research that the professor deems more appropriate to a less capable individual. Some apparently well meaning mentors may fall into this paternalistic trap, which may result in a department that attracts a diverse population of students, producing graduates segregated in their skills along lines of race or gender. This is not a service to those students of underrepresented identities whose career trajectories are oppressed by biased advice. It would be better to provide the same options to all of them and provide tailored information on how students can work toward their desired outcomes with encouragement. 

Evidence of microaggressions can also look like segregation or isolation, even if some would assert that it happened organically. People tend to gravitate toward where they feel welcome and respected. Most people can tell the difference between expressions of collegiality and respect are genuine and where they are put on like a mask. If there is a trend in any academic community of segregation or isolation along demographic axes, this may be evidence of a less welcoming environment or peer group. While no offense may be obvious, it is important in these cases to provide opportunities for genuine community building and to be vigilant for instances of harassment, discrimination, inequity, and more generally for disproportionate attrition.

Evidence of a culture of micro-aggressions also often looks like absence. While small number statistics at individual institutions are often used as an excuse for inequitable practices, nation-wide trends co-adding participation in physics academia tell a remarkably clear and consistent story. Women and people of color who are intelligent, capable, and formally trained in physics disproportionately prefer to work outside of academia. While their successes in industry, teaching, and entrepreneurship are to be commended, if their choices were coerced by an academic culture of bias, rejection, and in-group favoritism, then it is truly the high-energy physics that has lost with regard to access to their expertise and the discoveries they would have made if afforded the resources due to them.

\section{Special situation: COVID-19}

Since early 2020, the world has been immersed in the COVID-19 pandemic, bringing enormous challenges to our society and affecting all aspects of our life, including family and work. In the context of this paper, we can use the experience gained during these two years of the pandemic to evaluate the impact it has had on our wellness as individuals. In many cases, COVID-19 has increased the work-personal life imbalance discussed above, especially during the first year. Added duties during work hours like monitoring remote learning added extra stress and competition for attention. However, some changes like having more remote work and participation in conferences and meetings, gave us the possibility to better compensate our personal life with work tasks. Individuals did not have to spend time commuting to work or traveling abroad to participate in workshops or experimental activities that could be carried out remotely, allowing them to have more time for personal care and duties. In this section, we shall briefly discuss what we have learned and how we can use this experience to help individuals have a better balance between work and personal wellness in the future.

When the pandemic emerged, several countries implemented lockdowns, where most people were requested or forced to stay at home, working remotely, and taking care of their dependents (children, elders, and other individuals with needs). In most research fields, several studies have shown a clear decline in time devoted to research, which affected in a disproportional way different groups of researchers. Several studies have consistently shown that women scientists, especially those with young children, were the most affected. This has further increased the persistent and well-know gender gap in science, in particular, in our field, which will likely have an impact in the following years after the pandemic. 

The second pandemic year (2021) was different due to changing mitigation measures, adjustments in how societies handled COVID-19, and the development of multiple vaccines. With different approaches and at distinct times, depending on the country and the case peaks, individuals started to go back to their workplaces, students went back to schools (even if with interruptions), and society learned how to better handle the pandemic's challenges. 

In this sense, the initial impact of the decline of time devoted to research seems to have been alleviated. However, studies\footnote{\url{https://doi.org/10.1038/s41562-020-0921-y}} have shown a decrease on the rate of scientists starting new projects as a clear a consequence of the first pandemic year. This can be easily explained if we take into consideration that the pandemic and the associated social distancing measures halted many in-person interactions that might have facilitated the flow of new research ideas and collaborations. As expected, the fraction of decrease in new projects is also gender dependent. Due to the time necessary to mature and publish new ideas, we believe that the impact of the pandemic on the publication rate is not yet fully observable.

Other studies, based on surveys carried out in international collaborations\footnote{\url{https://cds.cern.ch/record/2752585/files/LHCb-PUB-2021-004.pdf?version=2
}}, have shown that the pandemic due to COVID-19 affected non-permanent researchers (graduate students and post-docs) and permanent researchers differently. In the survey, non-permanent researchers were generally uncomfortable working from home and wanted to work more frequently at the office/lab, while permanent researchers preferred working at home. This is not only correlated with the individual's career status, but also with the fact that in international collaborations, it is more common that non-permanent researchers are not in their home country. Additionally, permanent researchers had more agency to chose their working conditions than non-permanent researchers whose working conditions were often at the discretion of advisors. Regardless of the context of large collaborations, is is clear that the well-being of early career colleagues has deteriorated during the pandemic more than the more senior ones. This can be seen through decreased productivity, lack of motivation and focus on tasks, and signs of a decline in mental health since the start of the crisis. 

There were, nevertheless, some positive aspects of the new configurations that emerged during the pandemic. One of the obvious things, at least for researches from developing countries, was remote participation in meetings and conferences, which made access to these events more democratic. During the last two pandemic years, most events were held remotely, which allowed more geographically diverse participation since resources for travel were not a limiting factor. While remote meetings are more fatiguing, have time zone issues, and online discussions are much more difficult, more researches were able to attend. The development and wide availability of technology for remote meetings, have made it possible to have future events in a hybrid mode. This would allow people that can travel to attend in-person meetings, while those that would not be able to attend in-person, mainly due to lack of resources, could at least participate in most activities. Although remote and in-person participation do not entitle the same quality of communication, having hybrid meetings will contribute to diminish the existing gap between nations.


\section{Recommendations}
Personal wellness is of the highest importance and needs to be addressed as such. Structural and institutional barriers, high demands, and a lack of resources increase existing disparities and gaps in gender and underrepresented minorities. To have a more representative and equitable field, these issues must be urgently addressed.

Additionally, A lack of work-life balance can cause numerous negative effects on health both physically and mentally. Extended stress can cause burnout, exhaustion, and make people more susceptible to illness. Chronic stress can manifest in the body, causing tension, disrupting sleep, and causing serious health conditions that include depression, anxiety, diabetes, heart disease, gastrointestinal disease, high blood pressure, hyperthyroidism. Chronic stress from this imbalance can also take a toll on personal relationships, adding further stress. The stress can also lead to unhealthy coping mechanisms, including excessive alcohol, drugs, or food. While self-care and stress reduction techniques can sometimes lessen the effects of work/life imbalance, they are not a panacea. They themselves take time and do not treat the root causes of the stress. Additionally, they are not effective for all people.

The effects of chronic stress can be amplified for those that have pre-existing health conditions. Firstly, those with health conditions may often need additional time outside work to get treatment for their conditions. For example, someone who needs regular physical therapy could need several hours a week for regular treatment, while someone with depression might need weekly counseling. Work outside of nominal hours can even prevent treatment of conditions in some cases. Secondly, the chronic stress from work-life imbalance can take a physical toll on the body and mind, exacerbating existing conditions further.

It is critically important for the future and sustainability of our field to create a healthy and equitable work environment for members of our community at all levels and stages in their careers. The causes of work-life imbalances are far-reaching and systemic, requiring an approach that works to mitigate the problem at all levels: on an individual level, on an institutional level, on a project level, and on a funding agency level. The following sections provide recommendations to address these imbalances.

\subsection{Individuals}
On the individual level, all individuals should call out harmful rhetoric, including ideas around `lone geniuses', the need for unhealthy work schedules, and the idea that sacrifice of personal wellness demonstrates your commitment to science. Senior scientists can ensure that they are managing their time and the time of those in their group properly to allow for appropriate work-life balance. For example, when building schedules and setting deadlines, they should assume that members of their group will only be working during nominal work hours (no long days or weekends), include plans for time off, and account for other commitments that those in their group may have (e.g. leaving time for graduate student course work).

While it can be convenient to send correspondence during non-traditional work hours, it should be made clear that group members are not expected to respond outside of normal working hours. One way to do this is through the use of email signatures that clarify that a response is not expected outside of work hours. This is of particular importance when sending emails to those in different time zones. Even with clear expectations about responses, there is still pressure to respond to emails during off hours. Individuals should pause and consider if their emails can wait until normal work hours for the person that they are sending them to. Additionally, when setting meeting times, individuals should try to accommodate as many time zones as possible. One way to do this is through meetings that rotate their time every other meeting. This enables people to attend on weeks that are during nominal work hours in their timezone or to at reduce the number of meetings outside of nominal work hours.

Group leaders can also model taking time off and encourage their group members to do so as well. Ensure that those in your group know how to ask for time off and know about institutional policy and resources on diversity, health, leave, vacation, and wellness. 

\subsection{Department and Institution leadership}

Departments and institutions should have clear definitions of job responsibilities and ensure that they are funding all functions of the job. When only research and teaching are funded, scientists are often expected to perform other job functions on top of their normal work loads. For example, time spent on EDI efforts, outreach, and academic service should be accounted for in the funding and assessment criteria. Institutions and departments could include these tasks as different pay codes, and their assessments should weight work in these areas equally. They should establish awards and recognition for those who excel in this work.

Additionally, evaluation for admission or employment should be based on well thought out, prepared rubrics that include EDI, outreach, and service. These rubrics should be made publicly available and employees should know where to find them. The committees creating the rubrics should consider carefully and research if any of the criteria are biased in a way that would limit access by people who are traditionally underrepresented. Physics programs should also be structured to allow flexibility and finishing the degree in four years. For example, physics programs should allow freshmen who need to take a semester of calculus or who decide on physics as a major in their second year to still have the ability complete the degree in four years. 

Institutions and departments should also improve the pay and resources for graduate students and postdocs. They should be paid at the level of their respective skill levels, and relocation services that include moving expenses and partner placement should be provided. When stationed abroad, institutions need to ensure that employees and their families have health coverage. Institutions should also ensure that graduate student and postdoc pay is distributed in such a way that students can contribute to retirement savings (e.g. ensuring that they get a W-2 instead of only a 1098-T) and are not taxed for fellowship money they do not receive as pay. Institutions should also ensure that subsidized housing available to students and postdocs is also available for their partners and spouses.

Institutions should also have accessible, clear, robust, and flexible policies for parental leave, family leave, and vacation time ensure that they are guaranteed at all levels. For example, some institutions only allow graduate student off-time at the discretion of their research advisers, which sets up a dangerous dynamic. People at all career levels should be encouraged to take their leave and vacation time. One way to encourage taking vacation time could be to have set time periods where the institution is shut down and workers are encouraged to take time as in the pilot program at Fermilab.

Institutions should also have equitable policies on leaves of absence for health that include both physical and mental health. Additionally, they should ensure that they have equitable family leave policies for all genders. Institutions should offer training around mental health and wellness, how to make accommodations (including those for mental health), and how to cultivate supportive environments. Institutions should ensure that teachers and research mentors allow students time away during the day to tend to health concerns, including attending therapy. 

To reduce the power that individual advisers have over graduate students and postdocs, departments should implement group-based advising. This would reduce the power dynamics of the relationship and thus add a layer of protection for students and postdocs. Students ans postdocs should have at least two advisors at their institution or within their collaboration. Additionally, institutions should not require a letter of recommendation from PhD advisors for applications but instead allow applicants to choose their letter writers when letters are required. Institutions should also consider eliminating or putting less weight on letters of recommendation as they have been shown to exacerbate biases. Departments should also implement a system of faculty overseeing each other's advising through regular check-ins with a faculty member or faculty committee. Departments should also offer training for early career members on standards in the field.

Finally, institutions should develop reporting structures and cultivate an environment where concerns are appropriately heard and responded to. Harassers that are prominent scientists should receive the same response and consequences as those that are not. Responsive reporting structures ensure that barriers to personal wellness are both reported and addressed.

\subsection{Project Leadership}
On the project level, leaders and managers can also ensure that project schedules only assume nominal work hours and include vacation time. Additionally, project meetings should strive to accommodate as many time zones as possible and use rotating meeting times if necessary.

They should also ensure that all project members can access health care when they are on-site, including mental health care. For example, it is very difficult for students at CERN to get mental health care, so they often have to choose between health and their research at CERN.

Project policy should also have clear and transparent guidance on parental and family leave. For example, if a leader of a group goes on parental leave, there should be a clear process for covering their position while they are out and returning it to them when they return.

An effort should be undertaken to provide a more even distribution of the work load in collaborations, at the same time ensuring that all relevant tasks are given the same visibility and that high quality work is rewarded in the same way. This information should be conveyed by the collaborations as part of the job applications of their members. This should also be reflected in the choices of representatives of the collaborations to deliver talks at conferences, etc. Also, an effort has to be made to increase the cooperation within the collaboration, avoiding direct competition within individuals and their undesired effects in the duplication work and the mental health of the members. Collaborations should also train members in standards in the field and offer mentorship programs to ensure that postdocs and students have additional support and resources.

\subsection{Funding agencies}
Funding agencies can also help in these efforts through providing funding for other functions of research, including mentorship, EDI, and outreach efforts.  Often funding agencies expect these activities, but do not provide explicit budget for them, which encourages these activities to fall outside of nominal work tasks. Grant calls and assessments should include clear definitions of the tasks expected of PIs and provide grant funding for each of them. For example, funding agencies could require mentorship plans that include work-life balance and professional development for any personnel (students, post-docs, etc.) funded through the grants, EDI plans, and outreach plans.  Alternatively, agencies could provide specific grants and awards for EDI, outreach, and mentorship work. Agencies should also ensure that they pay those on their grant review panels for their time, and have professionals in EDI and education review mentorship, outreach, and EDI plans. Collecting PI and team demographic information on grant proposals could also be used to track the effectiveness of these measures and proposals by PIs.

While fellowship pay from funding agencies for postdocs and graduate students has traditionally been higher than that from institutions, it is still well below the market average. Pay for these positions needs to be increased and include relocation expenses.

Funding agencies could add additional pressure and accountability on institutions by requiring that any institution receiving funding from them has policies on vacation time, parental leave, family leave, and health leave for scientists at all levels,  especially for students and postdocs. Additionally, funding agencies should only fund institutions that prohibit confidentiality in harassment settlements to avoid enabling harassers to continue their behavior. Funding agencies could also provide funding for regular climate surveys (e.g. every 5 years). These could be used to capture statistics on hours per week spent working and how much vacation people can and do take and how these evolve over time.

\Acknowledgements{This manuscript has been authored by Fermi Research Alliance, LLC under Contract No. DE-AC02-07CH11359 with the U.S. Department of Energy, Office of Science, Office of High Energy Physics. T.L. acknowledges support from the NASA Postdoctoral Program at NASA Goddard Space Flight Center, administered by Oak Ridge Associated Universities.}













\end{document}